\documentclass[apjl]{emulateapj}
\usepackage{apjfonts}
\slugcomment{To appear in {\it The Astrophysical Journal Letters.}}
\lefthead{Tsuribe \& Omukai}
\righthead{Stellar Mass in Metal-Poor Environments}

\begin{document}

\title{
Physical Mechanism for the Intermediate Characteristic Stellar Mass in the
Extremely Metal-poor Environments
}

\author{
Toru Tsuribe\altaffilmark{1} and Kazuyuki Omukai\altaffilmark{2}
}

\altaffiltext{1}{Osaka University, Toyonaka, Osaka 560-0043, Japan; 
tsuribe@vega.ess.sci.osaka-u.ac.jp }

\altaffiltext{2}{
National Astronomical Observatory of Japan, Mitaka, Tokyo 181-8588, Japan; 
omukai@th.nao.ac.jp }


\begin{abstract}
If a significant fraction of metals is in dust,
star-forming cores with metallicity higher than a critical value 
$\sim 10^{-6}-10^{-5}Z_{\sun}$ are able to fragment 
by dust cooling, thereby producing low-mass cores.
Despite being above the critical metallicity,   
a metallicity range is found to exist around $10^{-5}-10^{-4}Z_{\sun}$ 
where low-mass fragmentation is prohibited.  
In this range, three-body H$_2$ formation starts
at low ($\sim$ 100K) temperature and thus the resulting heating causes 
a dramatic temperature jump, which makes the central part of 
the star-forming core transiently hydrostatic and thus 
highly spherical.
With little elongation, the core does not experience fragmentation 
in the subsequent dust-cooling phase.
The minimum fragmentation mass is set by the Jeans mass just before the 
H$_2$ formation heating, and its value can be as high as $\sim 10M_{\sun}$.
For metallicity higher than $\sim 10^{-4}Z_{\sun}$, 
H$_2$ formation is almost completed by the dust-surface reaction 
before the onset of the three-body reaction, and 
low-mass star formation becomes possible.
This mechanism might explain the higher characteristic mass 
of metal-poor stars than in the solar neighborhood  
presumed from the statistics of carbon-enhanced stars. 
\end{abstract}


\section{Introduction}
Early generations of stars affect the subsequent evolution of the universe 
in diverse ways, such as the cosmic reionization and metal-enrichment 
of the intergalactic medium.
In contrast to a number of theoretical attempts to understand
their nature in the last decade 
(e.g., Bromm \& Larson 2004; Ciardi \& Ferrara 2005), 
their observational confirmations are thus far extremely limited.
Among them, surveys of metal-poor stars in the Galactic halo 
provide unique tools to catch a glimpse of star formation in 
very low-metallicity environments in early galaxies 
(e.g., Beers \& Christlieb 2005).

In recent years, hundreds of extremely metal-poor (EMP) stars ([Fe/H]$<-3$) 
have been found in the halo.
A high fraction of them is noticeably enhanced with surface carbon abundance.
Such carbon-enhanced extremely metal-poor (CEMP) stars have 
attracted particular attention in constraining the stellar initial mass 
function (IMF) in metal-poor environments.
Especially, those with $s$-process element enhancement, CEMP-$s$ stars,
are considered to originate from secondary stars in binary systems.
They acquired the surface carbon during the AGB phase of 
intermediate-mass (several $M_{\sun}$) primary stars 
through the binary-mass transfer.
Although such carbon enhanced stars are present among 
population I stars as well 
(about 1 \%), the CEMP-$s$ stars make up far higher fraction 
(about 20 \%) of the EMP stars (e.g., Lucatello et al. 2005).
Since such a high fraction cannot be accounted for under the same IMF as 
in the solar vicinity, Lucatello et al. (2005) concluded that
the IMF is biased toward higher mass in the early Galaxy.
By using more sophisticated stellar evolution models,  
Komiya et al. (2007) reached similar conclusion.
Although there is apparent scarcity of stars in the ultra-metal poor (UMP) 
range ($-5<$[Fe/H]$<-4$) thus far (e.g., Salvadori et al. 2007), 
three stars has been discovered with 
[Fe/H]$<-4.8$ and all of them are carbon-enhanced 
(Christlieb et al. 2002; Frebel et al. 2005; Norris et al. 2007).
Suda et al. (2004) showed that their carbon enhancement can be explained by 
the binary-transfer hypothesis similar to that for the CEMP stars.
Extending the above argument to lower metallicity range, 
Tumlinson (2007) concluded that the IMF is even more skewed toward 
massive ones. 
It is quite natural to conceive that this trend persists 
also in the UMP range.
Moreover, the current lacking of UMP stars might suggest 
that even less low-mass stars are formed in this range.

Is there any possible mechanism that favors the formation 
of intermediate-mass stars in these metallicity ranges?
In metal-free gas, the first stars are expected to be very massive, 
more than $100M_{\sun}$, owing to inefficient fragmentation of 
dense cores whose mass scale is $\sim 1000M_{\sun}$ formed 
at $10^{4} {\rm cm}^{-3}$ (Bromm, Coppi, \& Larson 1999).
On the other hand, a slight amount of dust enables fragmentation 
at higher densities and then low-mass ($<1M_{\sun}$) core 
formation owing to the cooling by dust thermal emission 
(Schneider et al. 2002, 2003, 2006; Omukai et al. 2005; 
Tsuribe \& Omukai 2006 (paper I)). 
In this case, although the upper limit of stellar mass is not known, 
there is no particular reason favoring the formation of intermediate-mass 
stars.

Here, we propose a physical mechanism that prohibits low-mass star
formation in the metallicity range $-5 \la {\rm [M/H]}_{0} \la -4$ 
despite being above the critical metallicity ${\rm [M/H]}_{\rm cr}$ 
for dust-induced fragmentation. 
\footnote{We denote the metallicity in the star-forming gas 
as ${\rm [M/H]}_{0} \equiv {\rm log}(Z/Z_{\sun})$.}
This allows transition from intermediate-mass to low-mass star 
formation mode around the UMP to EMP range 
provided that a significant fraction of metals are already 
locked into dust by the time metallicity reaches this value.

\section{Effect of H$_2$ Formation Heating in Low-metallicity Gas}

Fragmentation properties of star-forming cores are largely determined 
by their thermal evolution.
In Figure 1, we show the temperature per unit hydrogen nuclei $T/\mu$  
$(\propto P/n_{\rm H})$
for metallicities ${\rm [M/H]_{0}}=-5.5, -4.5$, and -3.5 (solid lines).
Also those with $-5 \le {\rm [M/H]_{0}} \le -3.8$ 
are shown by dotted lines with an increment of 0.1.
These are calculated using a one-zone model as in Omukai et al. (2005),
except for the dust model, for which we use that produced in the 
pair-instability supernova of a metal-free $195M_{\sun}$ star 
(Schneider et al. 2006).
Dust from type II SN gives a similar result (see Schneider et al. 2006).
In the case of metallicity [M/H]$_{0} \sim$ -4.5, a sudden jump in 
$T/\mu$ is present 
around $n_{\rm H} \sim 10^8 {\rm cm^{-3}}$, which is caused 
by heat injection due
to three-body H$_2$ formation (Palla, Salpeter, \& Stahler 1983).
This temperature jump is observed only in the range 
$-4.9 \la {\rm [M/H]_{0}} \la -4.0$ (see dotted lines).
For lower metallicities (e.g., [M/H]$_{0} <-4.9$), the temperature is already 
high at the density where the three-body reaction begins and 
the consequence of heat injection is just a gradual increase in temperature.
On the other hand, for higher metallicities (e.g., [M/H]$_{0} >-4.0$), 
H$_2$ formation is almost complete, owing to dust surface reactions.
Thus, the three-body reaction plays minor role and the heat injection is small.
Only around [M/H]$_{0} \sim -4.5$ does heat injection by three-body reaction 
result in a dramatic temperature increase.

We expect that when the equation of state becomes stiff owing to the H$_2$
formation heating, 
a hydrostatic core forms and that it becomes spherical to a high degree.
Since an initial elongation of the core of $\ga O(1)$ is required
for fragmentation in the dust-cooling phase (paper I), 
this heating is expected to prevent the subsequent fragmentation. 
In the following section, we demonstrate that this is indeed the case 
by way of hydrodynamical calculations. 

\begin{figure}
\epsscale{1.23}
\plotone{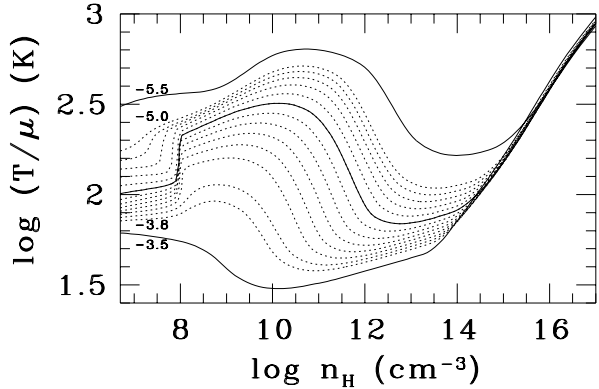}
\caption[dummy]{The temperature per unit hydrogen nuclei 
$T/\mu$ as a function of number density
for metal-poor star-forming cores with [M/H]$_{0}=-5.5, -4.5$, 
and $-3.5$ (solid).
Dust from the pair-instability supernova of $195M_{\sun}$
is assumed. 
Those for [M/H]$_{0}=-5...-3.8$ are also shown by dotted lines 
with an increment of $\Delta {\rm [M/H]_{0}}=0.1$. 
}
\label{fig1}
\end{figure}

\section{Hydrodynamical Calculations}

We study the evolution of cloud cores at high densities 
where the effects of the heating by three-body H$_2$ formation 
and cooling by dust are important by performing hydrodynamical calculations.
We use a barotropic equation of state (EOS) in Figure 1 for thermal evolution.
As in paper I, we choose the self-similar solution 
with $\gamma \simeq 1.1$ as an unperturbed initial state, 
to which both non-spherical density/velocity and
random-velocity perturbations are added in the same manner 
as in paper I.
The initial amplitude of elongation ${\cal{E}} \equiv
a/b-1$ is taken to be unity, where $a (b)$ is the short 
(long, respectively) core axis length.
In paper I, we started calculation just before the onset of the dust cooling 
without rotation.
Here to see the effect of H$_2$ formation heating, we start 
calculation at lower initial number density of $5\times 10^6{\rm cm^{-3}}$ 
at the center. 
As well as non-rotating cores, in this paper the cores with initial rigid 
rotation are considered.
Total mass of the cloud is 
$75 M_{\odot} (T_0/102{\rm K})^{3/2}$, where $T_0=P/(n_{\rm H}k_{\rm B})$ depends
on metallicity (Figure 1).
We used the smoothed particle hydrodynamics (SPH) based on the Godunov-type 
scheme (Tsuribe \& Inutsuka 1999; Inutsuka 2002) as in paper I, but 
with implementation of particle splitting by the condition that 
smoothing length is to be smaller than $1/14$ of the Jeans length.
In particle splitting, eight daughter particles are created on the vertices 
of a cube (Martel et al. 2006) in high density regions. 
The initial number of SPH particles is $N=5.2 \times 10^5$.
We remark that our method meets the resolution condition 
for physical fragmentation (e.g., Klein et al. 2004).

\subsection{Evolution of A Ultra Metal-Poor Core}
The evolution of the core with metallicity ${\rm [M/H]_{0}}=-4.5$
is shown in Figure 2 (a1-a4).
The initial core has angular velocity of $0.1 \sqrt{4 \pi G \rho_c}$, 
where $\rho_c$ is the initial density at the center.
As a result of the prominent H$_2$ formation heating at 
$n_{\rm H} \simeq 10^8 {\rm cm^{-3}}$, the pressure force exceeds gravity.
The collapse is almost halted along the short axis of the core 
while it continues along the long axis.
Consequently, the elongation of the core diminishes as
can be seen in Figure 3.
At this time, the core becomes almost hydrostatic and the accretion shock 
appears on the surface (Figure 2 (a2)).
The core continues to contract owing to accretion of the ambient medium. 
Although the core becomes elongated during the dust-cooling phase
($n_{\rm H} \ga 10^{11} {\rm cm}^{-3}$), 
for $n_{\rm H} \simeq 3\times 10^{14} {\rm cm}^{-3}$ the EOS becomes 
adiabatic and the growth of elongation slows down (see Figure 3).
During the dust-cooling phase, elongation grows at the disk formation epoch 
($n_{\rm H} \simeq 10^{14} {\rm cm}^{-3}$).
However, with a small value of elongation at the onset of dust cooling, 
the maximum elongation reaches at most $12$ (Figure 2 (a3); Figure 3). 
We calculated the cases with other values of initial angular velocity 
($0.01, 0.03, 0.3$, and $1.0 \sqrt{4 \pi G \rho_c}$) and found that
the maximum value of elongation does not exceed $20$ in any case.
Since this falls below the critical value for fragmentation 
$\sim 30$ (paper I), fragmentation does not occur.
We followed the evolution of the core until the early accretion stage and
confirmed that it does not fragment (see Figure 2 (a4)).
Therefore, the minimum fragmentation mass is set by the Jeans mass
before the H$_2$ formation heating, which is $\sim 10 M_{\sun}$.

\subsection{Evolution of Cores with Lower or Higher Metallicities}

Here we show that both the cores with ${\rm [M/H]_{0}}=-5.5$ and 
-3.5 fragment in the dust-cooling phase.
We illustrate the case of ${\rm [M/H]_{0}}=-5.5$ 
with initial angular velocity $0.1 \sqrt{4 \pi G \rho_c}$
in Figure 2 (b1-b4). 
For $n_{\rm H} \ga 10^{11} {\rm cm}^{-3}$ the temperature drops rapidly 
due to the dust cooling (Figure 1),
and the core becomes elongated to a filament (Figure 2 (b2,b3); Figure 3).
Although the temperature evolution becomes adiabatic 
after the core becomes optically thick, 
for $n_{\rm H} \ga 3 \times 10^{14}{\rm cm}^{-3}$, 
the elongation continues to increase owing to inertia.
At $n_{\rm H} \sim 10^{16}{\rm cm}^{-3}$, 
the elongation attains its maximum of ${\cal{E}}=49$, 
which is enough for the filament to fragment (paper I).
As seen in Figure 2 (b4), the filament fragments into five pieces 
and an additional five density peaks are still developing.
In the non-rotating case, the maximum elongation is $33$ 
(Figure 3), and two fragments are observed along with 
additional five growing density peaks.
That is, rotation enhances the elongation and promote the fragmentation 
of the cores.
This is because the elongation has longer time to grow 
during the collapse delayed by rotation.

We also calculated the case of higher metallicity ${\rm [M/H]_{0}}=-3.5$,
and found similar evolutionary features, although with larger 
maximum elongation of ${\cal{E}} = 94$ (Figure 3) even without rotation.
The core fragmented into at least seven pieces.

\section{Conclusion and Discussion}

The dust cooling produces low-mass fragments
of star-forming cores with metallicity as low as 
[M/H]$_{0}\la -5$, consistent with previous studies.
However, there is a pocket in the metallicity range
$-5 \la {\rm [M/H]_{0}} \la -4$, where the low-mass fragmentation 
is prohibited by sudden heat injection due to the three-body 
H$_2$ formation.
In this range, the fragmentation mass is $\ga 10M_{\sun}$. 
No low-mass ($< M_{\sun}$) core is formed, although fragmentation of 
this intermediate-mass core into a binary is not rejected.
Once the metallicity exceeds ${\rm [M/H]_{0}} \sim -4$, low-mass star formation
becomes possible by the dust cooling without hindrance by the three-body 
H$_2$ formation heating.
This provides a physical mechanism for the high characteristic mass 
($\ga$ several $M_{\sun}$; Tumlison 2007) of low-metallicity stars, 
which is suspected from the observed high frequency of low-metallicity 
carbon-enhanced stars.
In addition, the observed scarcity of ultra metal-poor stars, 
if confirmed, can be explained.
This also means that we need to modify the simplest scenario 
for the transition in star formation modes 
in the metal-poor gas; i.e., below 
a critical metallicity ${\rm [M/H]_{\rm cr}} (\simeq -6...-5)$
only massive stars are formed while low-mass stars are 
always formed in gas with metallicities above it
(e.g., Salvadori et al. 2006).

In this paper, we showed that H$_2$ formation on dust grains quenches 
the sudden heat injection by three-body H$_2$ formation 
at a metallicity [M/H]$_{0}\sim -4$
and enables low-mass star formation above this metallicity.
The range of metallicity $-4.9 \la {\rm [M/H]_{0}} \la -4.0$, 
where the sudden temperature jump appears, is slightly wider, but 
similar to the width of the observed gap in the MDF.
Although this coincidence is encouraging, it should not be taken at 
the face value since our calculation still contains uncertain parameters.
In particular, the nature of dust, e.g., gas-dust ratio, composition, and 
size distribution, in the early universe is very uncertain.
In the early universe, the dust is assumed to be produced in supernova events
and indeed a signature of dust has been observed in the extinction of 
high-z quasars and GRBs (Maiolino et al. 2004; Stratta et al. 2007).
On the other hand, in the nearby universe, there is no firm evidence 
for the dust production in supernova remnants (e.g., Bouchet et al. 2006).
Even if the dust production is efficient, a significant fraction of it 
might be destroyed by a reverse shock on the spot 
(Bianchi \& Schneider 2007; Nozawa et al. 2007).
In addition, the rate coefficient of three-body reaction rate is 
still uncertain, in particular, at low temperatures 
(Flower \& Harris 2007; Glover 2007).
Considering those uncertainties, we should admit that
although our prediction that the heating by three-body reaction 
prevents low-mass fragmentation in some metallicity range 
somewhere around [M/H]$_{0} \sim -4.5$ remains valid qualitatively, 
the exact metallicity range where this mechanism works would be altered.

Finally, we discuss effects of rotation.
Our result shows that the rotation accelerates the growth of 
elongation, thereby promoting the fragmentation.
This is consistent with the result by Clark, Glover, \& Klessen (2007),
who showed that, even in the primordial gas, where no fragmentation is
observed without rotation, rotating cores can fragment after 
the formation of rotation supported disks
although fragmentation is far less frequent than other cases.
In our cases, the core does not fragment in the metallicity range 
$-5 \la {\rm [M/H]}_{0} \la -4$ even with rotation.
Since parameters of the initial cores are not thoroughly surveyed here,
some cores might result in fragmentation in this metallicity range.
Even so, we expect that the number of such low-mass fragments is
smaller than in other range.
More studies on the effects of rotation on low-metallicity 
star formation are awaited.

\acknowledgments
We thank Simon Glover, Raffaella Schneider and an anonymous referee 
for comments improving the manuscript. 
Numerical computations were in part carried out on VPP5000
systems at the CfCA of NAOJ and the Altix 3700 at the Yukawa Insutitute 
(Kyoto).
This research was supported in part by Grants-in-Aid from 
MEXT of Japan (KO;18740117,18026008).



\begin{figure}
\epsscale{1.00}
\plotone{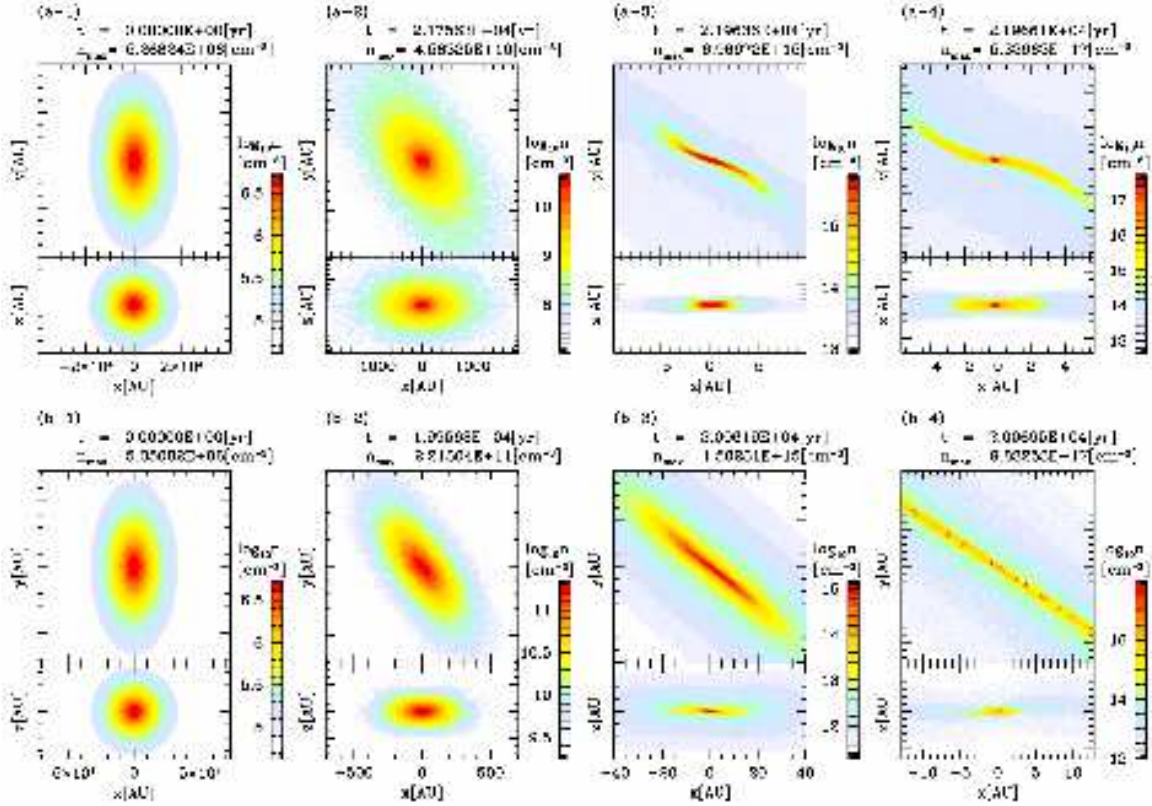}
\caption[dummy]{
Evolution of cores with 
(a) ${\rm [M/H]_{0}}=-4.5$ and  (b) $-5.5$.
The initial angular velocity is $0.1 \sqrt{4 \pi G \rho_c}$.
Four different stages (panels 1-4) are shown from left to right.
Density distributions in the $z=0$ plane (top), 
and the $y=0$ plane (bottom) are shown in each panel.
The panels 1 show the initial states, where the elongation is 
${\cal{E}}=1.0$.
The color scale denotes the density in the logarithmic scale.
The maximum number density and elapsed time are indicated on the top 
of each panel.
}\label{fig2}
\end{figure}

\begin{figure}
\epsscale{0.75}
\plotone{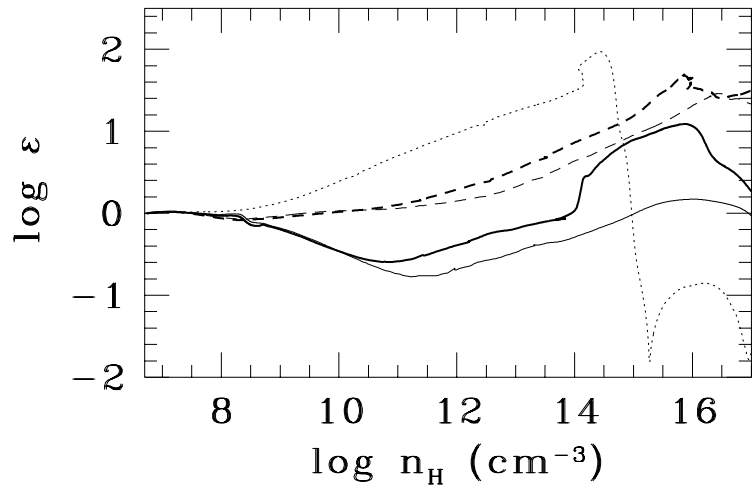}
\caption[dummy]{
Evolution of elongation ${\cal E}$ of the cloud cores 
with metallicity ${\rm [M/H]_{0}}=-5.5$ (dashed) and  $-4.5$ (solid).
Thick lines show the case with initial angular velocity 
$0.1 \sqrt{4 \pi G \rho_c}$, while thin lines show the cases
without rotation, respectively.
A non-rotating model with ${\rm [M/H]_{0}}=-3.5$ is also shown by 
the dotted line.
Elongation is calculated for the central region where the density 
is higher than $1/4$ of the maximum density.
The cores with ${\rm [M/H]_{0}}= -5.5$ and $-3.5$
fragment when their $\cal{E}$'s exceed the critical value $\sim 30$.
For the core with ${\rm [M/H]_{0}}= -4.5$, owing the decrease of elongation
by heating due to H$_2$ formation ($10^{8}-10^{12}  {\rm cm^{-3}}$), 
the elongation does not reach the critical value in the dust cooling phase.
}\label{fig3}
\end{figure}

\end{document}